\UseRawInputEncoding
\documentclass[%
 reprint,
 amsmath,amssymb,
 aps,prl,floatfix
]{revtex4-2}
 \usepackage[margin=1in]{geometry} 
\usepackage{amsmath,amsthm,amssymb,amsfonts}
\usepackage{float}
\usepackage{physics}
\usepackage{graphics}
\usepackage[utf8]{inputenc}
\usepackage{tikz}
\usetikzlibrary{positioning}
\usepackage{graphicx, subfigure}
\usepackage{float}
\usepackage{comment}
\usepackage{xcolor}
\usepackage{hyperref}
\usepackage{tcolorbox}
\usepackage{amsmath,amsthm,amssymb,amsfonts}
\usepackage{bbold}
\usepackage{physics}
\usepackage{graphicx}
\usepackage{dcolumn}
\usepackage{bm}
\usepackage{hyperref}
\usepackage{nicefrac}
\hypersetup{
	colorlinks = true,
	citecolor = {red}
	}
\usepackage[mathlines]{lineno}

\graphicspath{ {./images/} }

\newcommand{\vk}{\bm{k}}
\newcommand{\vp}{\bm{p}}

\newcommand{\VV}{\mathcal{V}}

\begin{document}
 
 \preprint{APS/123-QED}

\title{Momentum space entanglement of interacting fermions}

\author{Michael O. Flynn}
\email{moflynn@bu.edu}
\affiliation{%
Department of Physics, Boston University, 590 Commonwealth Avenue, Boston, Massachusetts 02215, USA
}
\author{Long-Hin Tang}
\affiliation{%
Department of Physics, Boston University, 590 Commonwealth Avenue, Boston, Massachusetts 02215, USA
}%
\author{Anushya Chandran}
\affiliation{%
Department of Physics, Boston University, 590 Commonwealth Avenue, Boston, Massachusetts 02215, USA
}%
\author{Chris R. Laumann}
\affiliation{%
Department of Physics, Boston University, 590 Commonwealth Avenue, Boston, Massachusetts 02215, USA
}%

\date{\today}

\begin{abstract}

Momentum space entanglement entropy probes quantum correlations in interacting fermionic phases. 
It is very sensitive to interactions, obeying volume-law scaling in general, while vanishing in the Fermi gas. 
We show that the R\'{e}nyi entropy in momentum space has a systematic expansion in terms of the phase space volume of the partition, which holds at all orders in perturbation theory.
This permits, for example, the controlled computation of the entropy of thin shells near the Fermi wavevector in isotropic Fermi liquids and BCS superconductors.
In the Fermi liquid, the thin shell entropy is a universal function of the quasiparticle residue.
In the superconductor, it reflects the formation of Cooper pairs. 
Momentum space R\'{e}nyi entropies are accessible in cold atomic and molecular gas experiments through a time-of-flight generalization of previously implemented measurement protocols.

\end{abstract}
\maketitle

Consider a many-body quantum system described by a wavefunction $|\psi\rangle$. 
For any partition of the system into regions $A$ and $\bar{A}$ the $n$th R\'{e}nyi entropy is
\begin{equation}\label{eq:Renyidef}
    S_{n}(A) = \frac{1}{1-n}\ln\text{Tr}\left[\rho_{A}^{n}\right]
\end{equation}
where $\rho_{A} = \text{Tr}_{\bar{A}}|\psi\rangle\langle\psi|$ is the reduced density matrix of subsystem $A$. 
Real space partitions have been extensively studied, as the scaling of $S_{n}(A)$ with the size of $A$ characterizes ground state properties in equilibrium as well as dynamical properties out of equilibrium~\cite{Calabrese:2009vm, review, Grover:2013wt,Laflorencie:2016vc,ETHreview, MBL}. 
The spectrum of eigenvalues of $\rho_A$ can also probe the physical excitation spectrum~\cite{LiHaldane} and the dynamical phase at non-zero temperature~\cite{MPlawstuff, ACCRLSid, RahulEntPhase}.

The real space R\'{e}nyi entropy has been measured in systems of ultracold bosonic atoms~\cite{islam_measuring_2015,kaufman_quantum_2016} and trapped ions~\cite{brydges_probing_random} using several protocols~\cite{daley_measuring_2012,brydges_probing_random}. Such measurements provide important experimental tests of quantum thermalization in isolated systems. 
Modified protocols have also been proposed for measuring real space entanglement in fermionic systems~\cite{pichler_thermal_2013,cornfeld_measuring_2019}.

\begin{figure}[tb]
\begin{center}
	\includegraphics[width=1\linewidth]{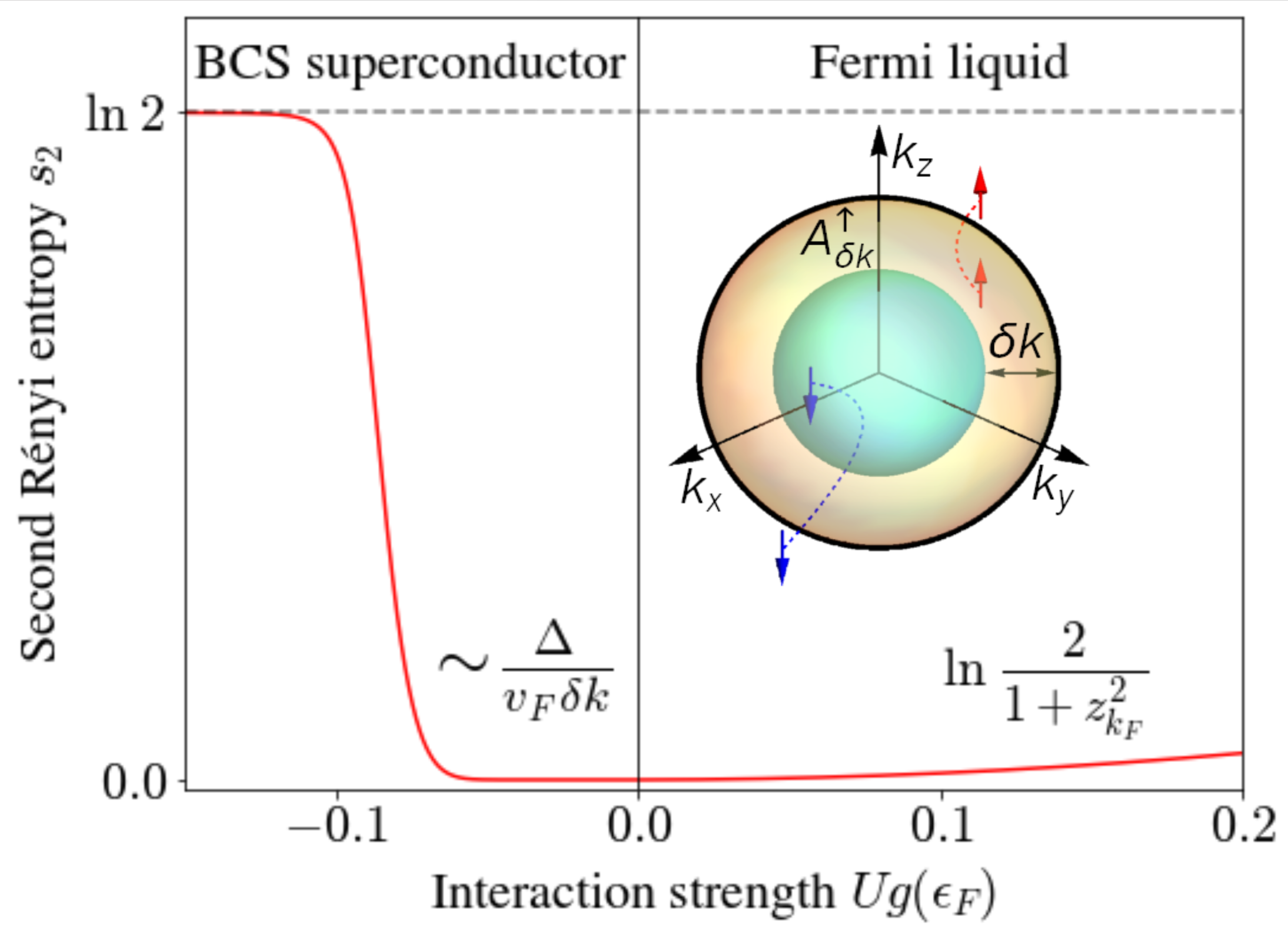}
	\caption{%
	The second R\'{e}nyi entropy per mode ${s_2(A^{\uparrow}_{\delta k})}$ for a spin-polarized, thin shell partition near the Fermi wavevector (inset). Here, $g(\epsilon_F)$ is the density of states at the Fermi energy, and the arrows in the inset indicate virtual processes that contribute to the entropy of the interacting ground state. 
	The entropy is controlled by the quasiparticle residue $z_{k_F}$ in the Fermi liquid, the gap $\Delta$ in the superconductor, and vanishes in the Fermi gas. 
	}
	\label{Fig:FELResultSummary}
	\end{center}
\end{figure}

For translation-invariant fermionic systems, it is natural to consider partitions of $|\psi\rangle$ in momentum rather than real space (real space cuts are discussed in Refs.~\cite{Klich,SwingleFermions}). 
Momentum space entanglement is extremely sensitive to interactions: in the ground state of the non-interacting Fermi gas, $S_{n}(A)=0$ for any momentum partition $A$. 
Generic interactions couple all momentum modes to one another, which implies that $S_{n}(A)\sim V|A|$, where $V$ is the volume of the system and $|A|$ is the $k$-space volume of $A$ (volume-law scaling)~\footnote{More precisely, we define $|A|$ to include spin/orbital degeneracies so that $V|A|$ is the number of modes in $A$}. 
The entropy per mode, $s_{n}(A)\equiv S_{n}(A)/V|A|$, thus characterizes the interacting system.

\begin{figure*}[t]
    \centering
    \includegraphics[width=\textwidth]{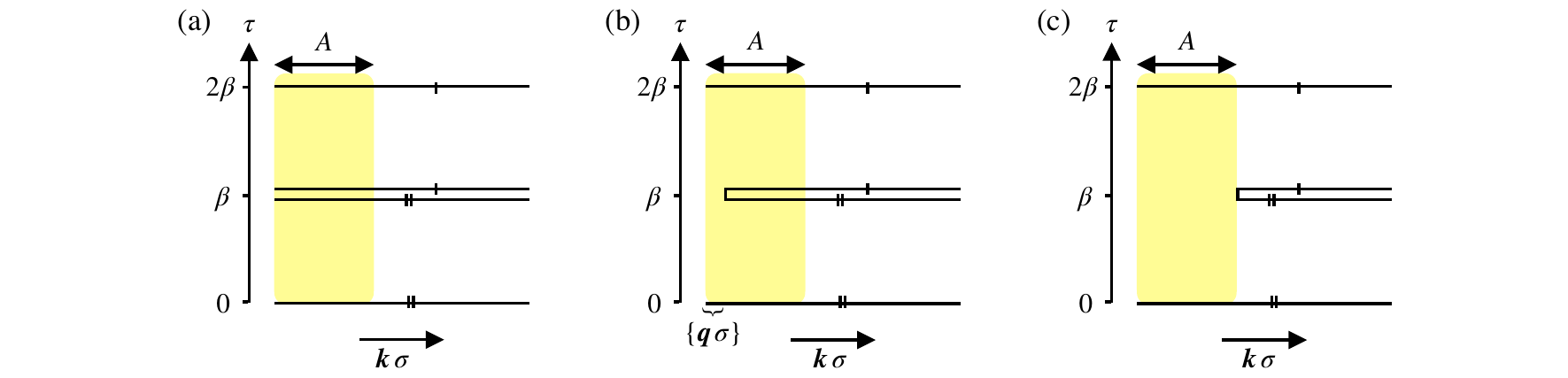}
    \caption{%
    The imaginary time manifold for pants in $\vk\sigma$-$\tau$ space with various waists $W$, arising in the computation of the R\'{e}nyi entropy: 
    (a) the normalization free energy $F^{(n)}(\emptyset)$ (``tubes''), 
    (b) the single mode R\'{e}nyi free energy $F^{(n)}(\{\bm{q}\sigma\})$  (``low-rise jeans'') and 
    (c) the $A$ R\'{e}nyi free energy, $F^{(n)}(A)$ (``pants''), all shown for $n=2$. 
    Vertical axis shows imaginary time extending from $0$ to $n \beta$, with boundaries cut and glued according to the markers. Horizontal axis is schematic representation of $\vk \sigma$ space.
    }
    \label{fig:manifolds}
\end{figure*}

In this manuscript, we compute $s_{n}(A)$ in the ground state of an isotropic Fermi system with short-range interactions (see Eq.~\eqref{fermiliquid}). 
This model realizes a Fermi liquid when the interactions are repulsive and a s-wave superconductor when they are attractive. 
In both phases, the lowest energy modes lie in thin shells near the nominal Fermi wavevector $k_{F}$, which is a natural regime to search for universal phenomena (see Fig. \ref{Fig:FELResultSummary} inset). 
Below, $A_{\delta k}$ denotes the set of modes with momenta in the range $\left[k_{F}-\delta k, k_{F}\right]$, and its spin-up (down) polarized counterparts are $A_{\delta k}^{\uparrow} (A_{\delta k}^{\downarrow})$.

We show that correlations between the different modes in $A_{\delta k}$ vanish as $\delta k \to 0$, such that the entropy $S_n(A_{\delta k})$ is simply the sum of the single mode entropies. 
We prove this result to all orders in interaction strength by relating the entropy to the free energy of interacting fermions on various pants-like manifolds (see Fig.~\ref{fig:manifolds}).

In the Fermi liquid, as the single mode entropy arbitrarily close to the Fermi surface is characterized by the quasi-particle residue $z_{k_{F}}$, the R\'{e}nyi entropies of thin shell cuts have \textit{universal} forms.
For example, the second R\'{e}nyi entropy is given by
\begin{equation}
\label{eq:universal_s2_fl}
    s_{2}(A_{\delta k})\underset{\delta k\to 0}{\longrightarrow} 2\ln\left[\frac{2}{1+z_{k_{F}}^{2}}\right] + \mathcal{O}(\delta k/k_F).
\end{equation}

In the s-wave superconductor, BCS theory predicts the presence of a superconducting gap $\Delta$ and Cooper pairing of fermions with opposite spin and momenta. 
Non-trivial momentum space partitions must trace out ``half'' of a Cooper pair; 
for partitions invariant under the transformation $\bm{k}\to -\bm{k}$, this requires that the partition is spin-polarized. 
For $A_{\delta k}^{\uparrow}$, the second R\'{e}nyi entropy is given by
\begin{equation}
\label{eq:superconductor_result}
    s_{2}(A_{\delta k}^{\uparrow})\sim \begin{cases}
\pi(1-2^{-1/2})\Delta/(v_{F}\delta k),\quad& v_{F}\delta k\gg\Delta\\
    \ln 2,\quad & v_{F}\delta k\ll\Delta\\
    \end{cases}
\end{equation}
where $v_F$ is the Fermi velocity. 
The saturation to the value $\ln 2$ reflects Cooper pairing throughout the thin shell.
These results for the Fermi liquid and superconductor are summarized in Fig.~\ref{Fig:FELResultSummary}.

Existing experimental protocols to measure real space entropy~\cite{daley_measuring_2012,brydges_probing_random} can be simply generalized to momentum space, as the underlying procedures do not prefer a particular single-particle basis prior to final measurements. 
We discuss the generalized schemes further below. 
Several groups have measured single-atom-resolved correlations in momentum space in various ultra-cold bosonic and fermionic systems in the last few years~\cite{Fang:2016,Hodgman:2017,Cayla:2018,Carcy:2019,Preiss:2019}, and have paved the way for the R\'{e}nyi entropy measurements that we propose. 

Momentum space entanglement has been previously studied in chiral and non-chiral fermionic systems. Momentum space partitions in the chiral quantum Hall setting are designed to probe the physics of a real space edge~\cite{LiHaldane, Sterdyniak:2012bs}, so their physics is quite different. In the non-chiral setting, various features have been reported in model studies in disordered systems~\cite{mondragon-shem_characterizing_2013, andrade_anderson_2014, ye_investigating_2017,lundgren_momentum-space_2019}, related spin chains~\cite{thomale_nonlocal_2010, lundgren_momentum-space_2014, Ibanez-Berganza:2016wi},  Luttinger liquids~\cite{dora_momentum-space_2016,Wei:2021wh}, Hubbard models~\cite{Anfossi:2008wt,ehlers_entanglement_2015} and field theories~\cite{balasubramanian_momentum-space_2012,hsu_momentum-space_2013}.

\paragraph{Fermi Liquids---}
Consider the following model of an isotropic Fermi liquid:
\begin{equation}
\begin{aligned}\label{fermiliquid}
    H&=H_0+H_1\\
    H_0&= \sum_{\bm{k},\sigma}\underbrace{(k^2/2m-\epsilon_{F})}_{\xi_{\bm{k}}}f^{\dagger}_{\bm{k}\sigma}f_{\bm{k}\sigma} \\
     H_{1}&=\frac{U}{V}\sum_{\bm{k}_1+\bm{p}_1=\bm{k}_2+\bm{p}_2}f^{\dagger}_{\bm{k}_2\uparrow}f^{\dagger}_{\bm{p}_2\downarrow}f_{\bm{p}_1\downarrow}f_{\bm{k}_1\uparrow}
\end{aligned}
\end{equation}
where $f_{\bm{k}\sigma}^{\dagger} \left(f_{\bm{k}\sigma}\right)$ are fermion creation (annihilation) operators with momentum $\bm{k}$ and spin $\sigma$, $U$ is the interaction strength, and $\epsilon_{F}$ is the Fermi energy. Throughout this manuscript, we take $\ket{\psi}$ to be the ground state.

Let us warm up by considering $A = \{\vk\sigma\}$ a  single spin polarized mode. In this case, number conservation dictates that $\rho_A$ is diagonal in the Fock basis with entries $\langle n_{\vk\sigma}\rangle$ and $1-\langle n_{\vk\sigma}\rangle$. The single mode R\'{e}nyi follows immediately,
\begin{align}
\label{eq:singlemode_renyi}
    S_n(\{\vk\sigma\}) &= \frac{1}{1-n}\ln \Big[\langle n_{\vk\sigma}\rangle^n + (1-\langle n_{\vk\sigma}\rangle)^n \Big]
\end{align}
If the mode lies near the Fermi surface, the occupation $\langle n_{\vk\sigma} \rangle \approx (1\pm z_{k_F})/2$ where we take $+/-$ for $\vk$ inside/outside the Fermi surface. 
Accordingly, the single mode entropy near the Fermi surface is an elementary function of the quasiparticle residue, $z_{k_F}$. 

In general, going beyond a single mode is analytically challenging in an interacting state. 
As an approximate approach, we start by neglecting all multimode connected correlations. 
This amounts to making a Gaussian approximation to the state $\rho_A$. 
For such states, it is well known that the R\'{e}nyi entropies follow from the one-body density matrix
\begin{align}
    \label{eq:onebodydm_def}
    G_{\vk'\sigma';\vk\sigma} = \langle f^\dagger_{\vk'\sigma'}f_{\vk\sigma} \rangle
\end{align}
restricted to the modes in $A$ \cite{peschel_calculation_2003},
\begin{align}
    S^G_n(A) & = \frac{1}{1-n}\tr \ln \Big[ G\vert_A^n + (\mathbb{1}-G\vert_A)^n \Big] \label{eq:SGaussian_Peschel}
\end{align}
For the Fermi liquid, momentum and spin conservation dictate that $G\vert_A$ is already diagonal for $\vk\sigma$-space cuts with eigenvalues given by the occupations $\langle n_{\vk\sigma}\rangle$. 
With reference to Eq.~\eqref{eq:singlemode_renyi}, we find that the Gaussian approximation predicts that the R\'{e}nyi entropy is simply the sum of the (exact) single mode entropies,
\begin{align}
\label{eq:gaussianRenyi_issum}
    S^G_n(A) &= \sum_{\vk\sigma\in A} S_n(\{\vk\sigma\})
\end{align}

For general partitions $A$, this approximation is uncontrolled. 
For example, if $A$ is the entire system, the true entropies vanish while Eq.~\eqref{eq:gaussianRenyi_issum} predicts an extensive positive value. 
On the other hand, $S^G_n$ is clearly exact for $A$ consisting of a single spin-polarized mode. 
More generally, short-range interactions in real-space lead to long-range interactions in $\vk\sigma$-space with an interaction strength between modes that scales inversely with the volume $V$, as in Eq.~\eqref{eq:model}.
Perturbatively, the associated connected correlations vanish for finite collections of modes; we thus expect that the Gaussian approximation is good for sufficiently small cuts $A$ in $\vk\sigma$-space. 

More precisely, in the appendix we show
\begin{align}
\label{eq:renyi_relationship}
    S_n(A_{\delta k}) - \sum_{\vk\sigma \in A_{\delta k}}S_n(\{\vk\sigma\}) \sim \mathcal{O}((\delta k/k_F)^2)
\end{align}
holds to all orders in perturbation theory in the coupling $U$ for thin shells $A_{\delta k}$.  
Formally, we obtain this result by relating the various R\'{e}nyi entropies in Eq.~\eqref{eq:renyi_relationship} to the free energy $F^{(n)}(W)$ of systems of interacting Grassmann fermions on pants-like manifolds in $\vk\sigma$-space and imaginary time  with varying waist regions $W$ (see Fig.~\ref{fig:manifolds}). 
Comparison of the diagrammatic expansion for $F^{(n)}$ on each of those manifolds allows us to show that the terms which contribute to Eq.~\eqref{eq:renyi_relationship} are indeed controlled by $\delta k$ at all orders. 

Putting Eqs.~\eqref{eq:singlemode_renyi}, \eqref{eq:gaussianRenyi_issum} and \eqref{eq:renyi_relationship} together for a cut $A_{\delta k}$ near the Fermi surface recovers the universal result quoted in the introduction, Eq.~\eqref{eq:universal_s2_fl}.

%%%%%%%%%%%%%%%%%%%%%%%%%%%%%%%%%%%%%%%%%%%
\paragraph{Superconductors---}
As the Cooper pairs in a s-wave superconductor are composed of fermions with opposite spin and momentum, it is natural to focus on spin-polarized partitions $A^\uparrow$ of momentum space. 

In this case, spin symmetry dictates that Eqs.~\eqref{eq:onebodydm_def},~\eqref{eq:SGaussian_Peschel},~\eqref{eq:gaussianRenyi_issum} still provide the Gaussian approximation to the Renyi entropy. 
In particular, $G\vert_{A^\uparrow}$ is diagonal and the anomalous correlator $\langle f_{\vk'\sigma'}f_{\vk\sigma} \rangle$ vanishes when restricted to $A^\uparrow$. 
Note that the Gaussian approximation with non-vanishing anomalous correlators in $A$ is different from that given in Eq.~\eqref{eq:SGaussian_Peschel} (see,  e.g.~\cite{peschel_calculation_2003}).

Furthermore, as the discussion around Eq.~\eqref{eq:renyi_relationship} suggests, the relationship between the single mode entropies and that of thin shells holds quite generally. In particular, the argument in the appendix readily extends to the s-wave superconductor after taking into consideration spontaneous symmetry breaking. 

In sum, the spin-polarized thin shell R\'{e}nyi entropies can be computed using Eqs.~\eqref{eq:SGaussian_Peschel}-\eqref{eq:renyi_relationship} in the s-wave superconductor.

Of course, in order to actually compute $S^G_n(A^\uparrow)$, one needs to know the occupation numbers of the $\{\vk\uparrow\}$ modes in the shell. BCS theory provides a self-consistent mean field approach to computing these occupations,
\begin{align}
    \langle f^\dagger_{\vk\uparrow} f_{\vk\uparrow} \rangle = \frac{1}{2}\bigg(1-\frac{\xi_{\vk}}{\sqrt{\xi_{\vk}^2+|\Delta|^2}}\bigg)
\end{align}
where $|\Delta|$ is the gap. Straightforward algebra produces,
\begin{equation}\label{eq:entkmode}
\begin{aligned}
    S_n(\vk\uparrow)&=\ln 2+\frac{1}{1-n}\ln\sum_{r=0}^{\left \lfloor{n/2}\right \rfloor }\binom{n}{2r}\bigg(\frac{\xi^2_{\bm{k}}}{\xi^2_{\bm{k}}+|\Delta|^2}\bigg)^r
\end{aligned}
\end{equation}

The scaling of ${s_n(A^{\uparrow}_{\delta k})}$ with ${|\Delta|}$ depends on its relative size with the energy scale of the thin shell. In the small gap limit ${|\Delta|\ll v_{F}\delta k}$, we can expand \eqref{eq:entkmode} in powers of ${|\Delta/\xi_{\vk} |}$ or ${|\xi_{\vk}/\Delta|}$ to obtain
\begin{equation}
     S_n(\bm{k}\uparrow)\approx
     \begin{cases}
     C(n)|\Delta \xi_{\bm{k}}^{-1}|^{2},\quad |\xi_{\bm{k}}|>|\Delta|\\
     \ln 2-n |\Delta \xi_{\bm{k}}^{-1}|^{-2}/2,\quad |\xi_{\bm{k}}|<|\Delta|
     \end{cases}
\end{equation}
where ${C(n)=\frac{2^{1-n}}{n-1}\sum_{r=0}^{\left \lfloor{n/2}\right \rfloor }\binom{n}{2r}r}$.
Summing up all contributions leads to the result in Eq.~\eqref{eq:superconductor_result}, which confirms the following simple intuition. 
When ${|\Delta|\gg v_{F}\delta k}$, all modes within the thin shell are strongly hybridized, resulting in the saturation of ${s_n(A^{\uparrow}_{\delta k})}$ to the maximal value ${\ln 2}$. 
On the other hand, when ${|\Delta|\ll v_{F}\delta k}$, ${s_n(A^{\uparrow}_{\delta k})}$ scales linearly in ${|\Delta|/ v_{F}\delta k}$, since only modes within a region ${|\Delta|}$ around the Fermi surface are strongly hybridized.

\paragraph{Free Dirac Transitions---}
Within BCS theory, the superconducting gap $\Delta$ exhibits an essential singularity at $Ug(\epsilon_{F}) =0$. 
The thin-shell momentum space entropy shown in Fig.~\ref{Fig:FELResultSummary} inherits this singularity. 
It is natural to conjecture that this is connected to the presence of a spectral gap for $U<0$, and that in analogy with real-space entanglement, momentum space entropy can exhibit non-analyticities in response to gap-inducing perturbations. 
We test this hypothesis by computing the second R\'{e}nyi entropy of spin polarized momentum balls around a Dirac point in $D$ spatial dimensions. 
On tuning the mass $m$, we find a generic non-analyticity
\begin{equation}
    s_{2}(A^{\uparrow}) \sim |m|^D\ln{|m|}
\end{equation}
in the free theory (see appendix). We leave the extension to the interacting critical theory to future work.

\paragraph{Measurement Protocol---}
A series of experiments with ultra-cold bosons have demonstrated that real space R\'{e}nyi entropy can be measured by preparing copies of a quantum state and interfering them appropriately~\cite{islam_measuring_2015,kaufman_quantum_2016}. Here we briefly review this protocol, which has been extended theoretically to fermionic systems as well~\cite{cornfeld_measuring_2019,pichler_thermal_2013}, and generalize it to momentum space. 

Begin with two identical copies of a quantum state in a pair of optical lattices; typically these are prepared by independent but identical time evolution in each copy. A beam splitter then interferes the two copies by freezing each of their dynamics and allowing for tunneling between them using an optical superlattice. This operation maps fermions in the first $(a^{\dagger})$ and second $(b^{\dagger})$ copies as:
\begin{equation}
    a_{i,\sigma}^{\dagger}\rightarrow\frac{a_{i,\sigma}^{\dagger}+b_{i,\sigma}^{\dagger}}{\sqrt{2}};\hspace{12pt} b_{i,\sigma}^{\dagger}\rightarrow\frac{b_{i,\sigma}^{\dagger}-a_{i,\sigma}^{\dagger}}{\sqrt{2}} 
\end{equation}
Microscopy techniques are then used to measure site and spin-resolved particle densities, from which the second R\'{e}nyi entropy of an arbitrary real space partition is calculated~\footnote{See \cite{pichler_thermal_2013} for additional comments regarding ordering ambiguities and \cite{cornfeld_measuring_2019} for the generalization to higher R\'{e}nyi entropies.}.

Prior to measuring particle densities in real space, this protocol does not privilege any particular single-particle basis; it is the measurement basis which determines the partitions that can be accessed. Replacing real-space microscopy with a time-of-flight (TOF) single-atom-resolved measurement~\cite{Fang:2016,Hodgman:2017,Cayla:2018,Carcy:2019,Preiss:2019} enables the computation of momentum space R\'{e}nyi entropies in near-term experiments. In TOF, the atoms are released from the optical lattice and absorption imaging is used to reconstruct the initial momenta~\cite{bloch_many-body_2008}.

\paragraph{Discussion---}
The momentum space entanglement $S_n(A)$ in the ground state of interacting fermionic systems permits a systematic expansion in the phase space volume of $A$. 
The leading contribution for small $A$ is given by the sum of the single mode entropies -- this provides an explicit relationship between the R\'{e}nyi entropy and certain low energy properties of the phases for thin shells near the Fermi wavevector.

In the Fermi liquid, the leading term in the shell width $\delta k$ only depends on the quasiparticle residue, $z_{k_F}$. 
An interesting avenue for future work would be to compute the $O(\delta k^2)$ contribution, where we expect the Landau parameters to play a role as they reflect the correlations between $k$-modes. 

Similarly, the one dimensional interacting Fermi system realizes a Luttinger liquid where $z_{k_F} = 0$. 
Despite the more dramatic reorganization of the ground state from the non-interacting Fermi sea, we nonetheless expect $s_2(A_{\delta k}) =  2 \ln 2  + O(\delta k)$. 
It would be interesting to check this in a direct multi-mode calculation, building on Ref.~\cite{Ibanez-Berganza:2016wi}.

Although we have presented the results in the context of short-ranged interactions, the diagrammatic proof for the phase space expansion is much more general. 
In particular, we believe it can be generalized to Coulomb interacting systems and to unconventional superconductors.

\begin{acknowledgments}
The authors are grateful to C. Chamon, G. Goldstein, D. Long, and A. Nahum for stimulating discussions. 
This work was supported by the Air Force Office of Scientific Research through grant No. FA9550-16-1-0334 (M.O.F.) and by the National Science Foundation through the awards DMR-1752759 (L.-H.T and A.C.) and PHY-1752727 (C.R.L.).
\end{acknowledgments}

\bibliographystyle{unsrt}
\bibliography{citations}

\newpage

\begin{widetext}
\appendix

\section{Diagrammatic Treatment of Thin Shell R\'{e}nyi Entropy in Interacting Fermi Systems}
\label{App:FL}

For specificity, we consider an isotropic, weakly interacting Fermi system with short-range interactions, 
\begin{align}\label{eq:model}
    H &= \sum_{\vk,\sigma}\xi_{\vk}f_{\vk\sigma}^{\dagger}f_{\vk\sigma} + \frac{U}{V}\sum_{\vk_{1},\vp_{1},\vk_{2},\vp_{2}}\delta_{\vk_1+\vp_1 ,\vk_2 + \vp_2}f_{\vk_{2}\uparrow}^{\dagger}f_{\vp_{2}\downarrow}^{\dagger}f_{\vp_{1}\downarrow}f_{\vk_{1}\uparrow}
\end{align}
We will briefly discuss generalizations at the end. Our goal is to show that for a thin shell $A_{\delta k}$ of thickness $\delta k$ near the Fermi surface at $k_F$,
\begin{align}
\label{eq:renyi_conj}
    S_n(A_{\delta k}) - \sum_{\vk \sigma \in A_{\delta k}}S_n(\{\vk\sigma\}) \sim O((\delta k/k_F)^2)
\end{align}
That is, the full R\'{e}nyi entropy in the thin shell is the sum of the single mode R\'{e}nyi entropies to leading order in $\delta k$. 
In order to do this, we will express the relevant R\'{e}nyi entropies in terms of the free energies of interacting Grassmann fields on several different imaginary time geometries.
We will then prove Eq.~\eqref{eq:renyi_conj} by showing that the diagrams that contribute to the right hand side are suppressed by phase space factors at all orders in the interaction. 

Although we are ultimately interested in entanglement in the ground state of $H$, it is convenient to work at finite temperature, $\rho = e^{-\beta H}$, and take $\beta \to \infty$ at the end. The R\'{e}nyi entropy of a subset $A$ of momentum modes of the thermal state is given by 
\begin{align}
\label{eq:renyi_abstract}
    (1-n) S_n(A) &=  \ln \tr \rho_A^n - \ln \left(\tr \rho\right)^n \\
    \label{eq:free_energy_def}
    &\equiv -\beta (F^{(n)}(A) - F^{(n)}(\emptyset))
\end{align}
where $\rho_A = \tr_{\bar{A}} \rho$ is the (unnormalized) reduced density matrix in $A$, and the second term accounts for the normalization of $\rho$. 

Each of the two terms defined in Eq.~\eqref{eq:free_energy_def} can be understood as the free energy $F^{(n)}(W)$ of a system of Grassmann fields on an appropriate ``pants-like'' imaginary time manifold with waist $W$ in $\vk\sigma$-space (see Fig.~\ref{fig:manifolds}). 
Each of the $n$ legs has circumference $\beta$ while the waist region has circumference $n\beta$. 

Let us warm up by recalling the imaginary time representation of the usual free energy $F$ of $H$ at temperature $\beta^{-1}$:
\begin{align}
    Z &= e^{-\beta F} =  \tr e^{-\beta H} = \int D\bar{\psi}D\psi e^{-S} \\
    S &= \int_0^\beta d\tau 
    \sum_{\vk\sigma}\bar{\psi}_{\vk\sigma} \left( \partial_\tau + \xi_{\vk} \right)\psi_{\vk\sigma}
    + \underbrace{\frac{U}{V}\sum_{\vk_{1},\vp_{1},\vk_{2},\vp_{2}}\delta_{\vk_1+\vp_1 ,\vk_2 + \vp_2}\bar{\psi}_{\vk_{2}\uparrow}\bar{\psi}_{\vp_{2}\downarrow}\psi_{\vp_{1}\downarrow}\psi_{\vk_{1}\uparrow}}_{\VV}
\end{align}
The Grassmann fields satisfy anti-periodic boundary conditions on a circle of imaginary time of extent $\beta$ for all modes $\vk \sigma$:
\begin{align}
    \psi(0) &= - \psi(\beta) & \bar{\psi}(0) = - \bar{\psi}(\beta)
\end{align}
This leads to the non-interacting imaginary-time ordered propagators
\begin{align}
    G(\vk'\sigma'\tau_2;\vk\sigma\tau_1) &= \langle T_\tau \psi_{\vk'\sigma'}(\tau_2) \bar{\psi}_{\vk\sigma}(\tau_1)\rangle =  \delta_{\vk\vk'}\delta_{\sigma\sigma'} G_{\vk\sigma}(\tau_2, \tau_1) \nonumber\\
    G_{\vk\sigma}(\tau_2,  \tau_1) &= e^{-\xi_{\vk} (\tau_2-\tau_1)} (\theta(\tau_2-\tau_1) - f(\beta\xi_{\vk}))
\end{align}
where $f(x) = 1/(e^x+1)$ is the Fermi-Dirac distribution and $\theta$ is the step function. 

The diagrammatic expansion of $F$ is given by a sum of connected diagrams according to the linked cluster theorem,
\begin{align}
\label{eq:linkedcluster_F}
    - \beta F &= -\beta F_{0} + \sum_{l=1}^\infty \frac{(-1)^l}{l!} \left\langle T_\tau  \int_0^\beta d\tau_1 \cdots \int_0^\beta d\tau_l \VV(\tau_1)\cdots \VV(\tau_l) \right\rangle_0^c
\end{align}
We represent this expansion graphically in terms of diagrams in $\vk\sigma$-$\tau$ space built out of vertical propagator lines (at fixed $\vk\sigma$) and horizontal dashed interaction vertices (at fixed $\tau$, see Fig.~\ref{fig:diagrams}). 
A diagram corresponds to a particular Wick contraction of the underlying Grassmann fields and it is connected so long as each vertex $\VV(\tau_i)$ can be connected to any other by following the propagator lines. 
Momentum conservation implies that each connected diagram makes an extensive contribution to $F$ after integration over the loop momenta. 
Thus, $F \propto V$ as expected.

\begin{figure}[tb]
    \centering
    \includegraphics[width=\textwidth]{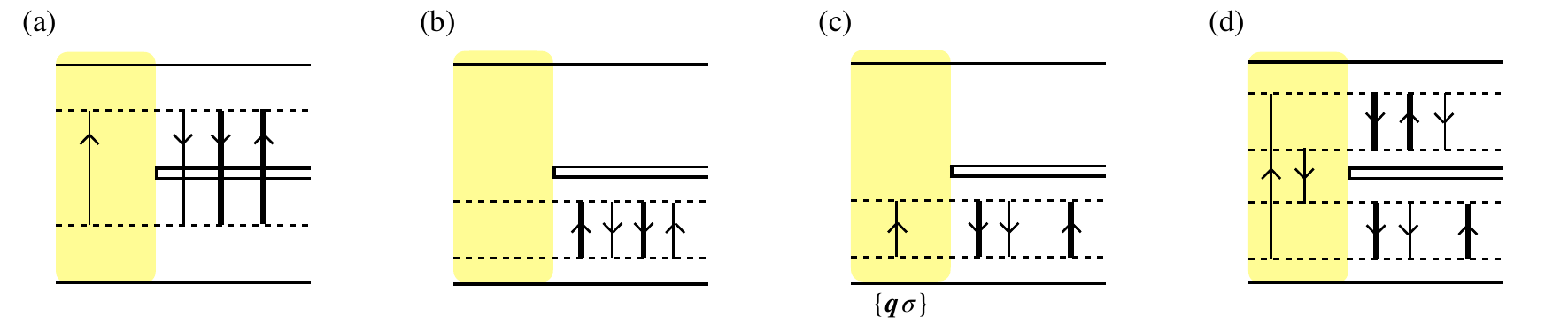}
    \caption{%
    Representative connected diagrams in the expansion of R\'{e}nyi free energies.  Interaction insertions $\VV(\tau)$ are indicated by horizontal dashed lines at time $\tau$. Vertical propagators are at fixed momenta (horizontal axis). Arrows show flow of fermion number and momentum, and spin up (down) is indicated by thin (thick) lines. 
    All diagrams are drawn on the `pants' geometry for reference but can be equally drawn on other backgrounds.
    (a) A second order connected diagram with one propagator in region $A$. Since several of the propagators extend between legs, the value is zero. 
    (b) A second order connected diagram with no propagators in $A$. This takes a finite value identically across all geometries.
    (c) A second order diagram with one propagator in $A$ which contributes to $S_2(A)$ but cancels in Eq.~\eqref{eq:renyi_conj}.
    (d) A 4th order diagram with two propagators in $A$. This contributes to the right hand side of Eq.~\eqref{eq:renyi_conj}. 
    }
    \label{fig:diagrams}
\end{figure}

The extension of this treatment to the geometries of Fig.~\ref{fig:manifolds} requires two modifications -- we extend the range of $\tau$ to $[0, n \beta]$ and modify the boundary conditions on the Grassmann fields depending on whether they are in the waist ($W$) or legs ($\bar{W}$) region of $\vk \sigma$ space. 
This leads to the following propagator
\begin{align}
\label{eq:waist_propagator}
    G^W_{\vk\sigma}(\tau_2, \tau_1) = \left\{ \begin{array}{ll}
    e^{-\xi_{\vk} (\tau_2 - \tau_1)}(\theta(\tau_2-\tau_1) - f(n \beta \xi_{\vk})) &\quad \vk\sigma \in W \\
    e^{-\xi_{\vk} (\tau_2 - \tau_1)}(\theta(\tau_2-\tau_1) - f(\beta \xi_{\vk}))\mathbb{I}\left[\tau_1, \tau_2 \textrm{ in same leg}\right] &\quad \vk\sigma \in \bar{W}
    \end{array} \right. 
\end{align}
Here, $\mathbb{I}$ is an indicator function taking values $0$ or $1$. The dependence on the absolute times rather than just their difference reflects the breaking of time translation invariance by the stitching of the legs.

The free energy $F^{(n)}(W)$ again can be computed from the linked cluster theorem,
\begin{align}
       - \beta F^{(n)}(W) &= -\beta F^{(n)}_{0}(W) + \sum_{l=1}^\infty \frac{(-1)^l}{l!} \left\langle T_\tau  \int_0^{n\beta} d\tau_1 \cdots \int_0^{n\beta} d\tau_l \VV(\tau_1)\cdots \VV(\tau_l) \right\rangle_{0,W}^c
\end{align}
The construction of connected diagrams is identical across manifolds with different waists $W$ -- their values differ only through the difference in the propagator.
For example, Figs.~\ref{fig:diagrams}(a,b,c) all represent contributions to the following term,
\begin{multline}
    V U^2 \int_{W \cup \bar{W}}\prod_{i=1}^4 \frac{d^3k_i}{(2\pi)^3}(2\pi)^3\delta^3(\vk_1 - \vk_2 + \vk_3 - \vk_4) \int_0^{n\beta}d\tau_1d\tau_2\, 
    \times \\
    G^W_{\vk_1\uparrow}(\tau_1,\tau_2) G^W_{\vk_2\uparrow}(\tau_2,\tau_1) G^W_{\vk_3\downarrow}(\tau_1,\tau_2) G^W_{\vk_4\downarrow}(\tau_2,\tau_1) \nonumber
\end{multline}
Thus, in this representation, we can compare the free energies on different manifolds diagram by diagram at any order simply be comparing which propagators lie in the relevant waists.

Now let us return to our conjecture Eq.~\eqref{eq:renyi_conj}. 
In terms of free energies, this is equivalent to
\begin{align}
\label{eq:free_energy_diff}
    \beta(F^{(n)}(A_{\delta k}) - F^{(n)}(\emptyset)) - \beta\sum_{\vk\sigma \in A_{\delta k}}(F^{(n)}(\{\vk\sigma\}) - F^{(n)}(\emptyset)) \sim O((\delta k/k_F)^2)
\end{align}
It is straightforward to check that the non-interacting contributions $-\beta F^{(n)}_0$ to the free energy are identical across all geometries as $\beta \to \infty$ and thus cancel in Eq.~\eqref{eq:free_energy_diff}. 
Let us organize the interaction contributions according to the number $m$ of distinct $\vk\sigma$ in $A_{\delta k}$ which some propagator inside the associated connected diagram carries. 

If $m=0$, all propagators in the diagram are leg propagators independent of the underlying geometry (see Fig.~\ref{fig:diagrams}b). Such diagrams clearly cancel within each bracket in the LHS of Eq.~\eqref{eq:free_energy_diff}.

For $m=1$, exactly one mode in $A_{\delta k}$, say $\bm{q} \sigma$, is active in the diagram (eg. Fig.~\ref{fig:diagrams}a,c).  Such a diagram takes different values in $F^{(n)}(A_{\delta k})$ and  in $F^{(n)}(\emptyset)$ and thus contributes to $S_n(A_{\delta k})$. However, the same contribution appears in $F^{(n)}(\{\bm{q}\sigma\}) - F^{(n)}(\emptyset)$ and thus cancels in Eq.~\eqref{eq:free_energy_diff}.

When integrated over $\bm{q} \sigma \in A_{\delta k}$, these $m=1$ diagrams produce the  leading in $\delta k$ dependence of the ``volume-law'' scaling of $S_n(A_{\delta k}) \sim V (k_F^2 \delta k)$.

Diagrams with $m\ge 2$ contribute to $S_n(A_{\delta k})$ as well as several $S_n(\{\vk \sigma\})$ (see Fig.~\ref{fig:diagrams}d). These contributions need not  cancel. However, on integrating over all momenta $\{\vk\sigma\}$ in the diagram, the requirement that $m$ of them lie in $A_{\delta k}$ reduces the phase space volume by $ \left(\delta k / k_F\right)^m$ relative to the free integration over all loop momenta. Thus, the leading contributions to the RHS of Eq.~\eqref{eq:free_energy_diff} enter at $(\delta k/k_F)^2$.

Let us close with a few comments on the generality of the above formal argument.
First, it is clear that the argument made very little use of the actual structure of the interaction and should readily go through for other sufficiently short-range momentum-conserving interactions.
Second, the choice of the non-interacting part of $H$ must respect the choice of cut $A$ -- more precisely, the non-interacting propagator $G_0$ must not connect $\vk\sigma \in A$ to $\vk \sigma' \in \bar{A}$ in order for Eq.~\eqref{eq:waist_propagator} to hold. 
The result Eq.~\eqref{eq:renyi_conj} nonetheless holds for spin/orbital-polarized cuts in a system with quadratic off-diagonal terms (eg. $A^{\uparrow}$ for a system in a transverse Zeeman field or for a mean-field superconductor). 
One simply treats the off-diagonal terms as part of $\mathcal{V}$ and introduces corresponding vertices with degree 2 in the diagrams.
While this is not usually the most efficient approach to computing Renyi's (or other physical properties) in such a system, it serves for the formal argument.
%Moreover, the Dyson equation for the propagator with the degree 2 vertices guarantees that the perturbative approach reproduces more direct treatments of the quadratic part of the theory if summed to all orders.
Finally, so long as there is a perturbative resummation scheme to obtain spontaneous symmetry breaking states (as in the BCS superconductor), we expect the result Eq.~\eqref{eq:renyi_conj} to hold. 
In this case, one should consider the diagrammatic calculation in the presence of a weak applied symmetry breaking field in order to break the symmetry explicitly. The weak field is taken to zero after the thermodynamic limit. 
This serves to select the particular symmetry breaking direction for the relevant self-energies. 

%%%%%%%%%%%%%%%%%%%%%%%%%%%%%%%%%%%%%%%%%%%%%%

\section{R\'{e}nyi entropies of Dirac Hamiltonians}

\label{App:Dirac}
In the main text, we analyzed the momentum space entropy of the fermion model \eqref{eq:model} for both attractive and repulsive couplings. By analogy with real-space entanglement, one might anticipate that momentum space entropy can exhibit a non-analytic response to a gap-inducing perturbation. In the case of the Fermi liquid to superconductor transition, this postulate is somewhat obscured by the fact that the Fermi liquid is gapless over an entire surface in momentum space. It is conceptually simpler and analytically better controlled to test this postulate in the context of fermionic models which are gapless at isolated points in momentum space.

Here we consider a simple model of fermions which is gapless at a Dirac point, and study how suitable entanglement properties of the model are modified by gap-inducing perturbations. These calculations confirm that the second R\'{e}nyi entropy of orbital-momentum cuts which enclose the Dirac point are non-analytic functions of a mass gap $m$.

The Dirac Hamiltonian in $D=2$ spatial dimensions reads
\begin{equation}
\begin{aligned}
    h(\bm{k})=k_x\sigma_{x}+k_y\sigma_y+m\sigma_{z}&= \left(k_{x},k_{y},m\right)\cdot\bm{\sigma}\\
    &= \sqrt{k^{2}+m^{2}}\left(\hat{n}\cdot\bm{\sigma}\right)
    \end{aligned}
\end{equation}

\noindent
where $\sigma_{i}$ are the Pauli matrices and $(k_{x},k_{y})$ labels momentum modes around the Dirac point at the origin. In the final equality, we introduced a unit vector $\hat{n}\equiv\left(\sin\theta\cos\phi,\sin\theta\sin\phi,\cos\theta\right)\propto\left(k_{x},k_{y},m\right)$. The eigenstates of $h(\bm{k})$ can be written as

\begin{equation}\label{eq:eigenstates}
|\epsilon_{+}(\bm{k})\rangle =
    \begin{pmatrix}
    \cos\left(\theta/2\right)\\ \sin\left(\theta/2\right)e^{i\phi} 
    \end{pmatrix},\hspace{12pt} |\epsilon_{-}(\bm{k})\rangle =
    \begin{pmatrix}
    \sin\left(\theta/2\right)\\ -\cos\left(\theta/2\right)e^{i\phi} 
    \end{pmatrix}
\end{equation}

\noindent
with associated eigenvalues $\epsilon_{\pm}(\bm{k}) = \pm\sqrt{k^{2}+m^{2}}$. In the ground state, negative energy states are populated and the ground state density matrix factorizes over momenta:

\begin{equation}
\begin{aligned}
\rho &= \bigotimes_{\bm{k}}\rho_{\bm{k}},\\ \rho_{\bm{k}} &\equiv \ket{\epsilon_{-}(\bm{k})}\bra{\epsilon_{-}(\bm{k})}
\end{aligned}
\end{equation}

It follows that for any subset of momenta $A$, $S_{n}(A) = 0$. Pure momentum-space partitions are therefore insufficient to obtain non-trivial entanglement: we must also trace out an orbital for each $\bm{k}$. This orbital is parameterized by a unit vector $\hat{a} \equiv \left(\sin\alpha\cos\beta,\sin\alpha\sin\beta,\cos\alpha\right)$, and we choose to trace out states that satisfy $\hat{a}\cdot\bm{\sigma}=-1$. The resulting density matrix, $\rho_{\bm{k}}^{\hat{a}+}$, is diagonal with eigenvalues given by

\begin{equation}
\begin{aligned}
\rho_{\bm{k}}^{\hat{a}+} &= \text{diag}\left\{\lambda(\bm{k}),1-\lambda(\bm{k})\right\} \\
\lambda(\bm{k}) &= \frac{1+\langle\hat{a}\cdot\bm{\sigma}\rangle}{2} \\
\langle\bm{\sigma}\rangle &= -\frac{1}{\epsilon_{k}}\left(k_{x},k_{y},m\right)
\end{aligned}
\end{equation}

For a collection of momenta $A$, the R\'{e}nyi entropies of fermions polarized along $\hat{a}$, $S_{n}(A^{\hat{a}+})$, is reduced to sums over independent modes. Specializing to $n=2$,

\begin{equation}\label{eq:EntSum}
    S_{2}(A^{\hat{a}+}) = \sum_{\bm{k}\in A}s_{2}^{\hat{a}+}(\bm{k})\to \frac{V}{\left(2\pi\right)^{2}}\int_{\bm{k}\in A}s_{2}^{\hat{a}+}(\bm{k})d^2k
\end{equation}

\noindent
where $s_{2}^{\hat{a}+}(\bm{k}) = -\text{Tr}\left(\rho_{\bm{k}}^{\hat{a}+}\right)^{2}$. A natural choice for the set of momenta $A$ is a circle of radius $k_{A}$ in momentum space centered on the Dirac point. The gap $m$ provides a scale which separates the momenta into regions $A_{m}=\left\{\bm{k} \hspace{6pt}| 0<|\bm{k}|<k_{A}\right\}$ and $\overline{A}_{m}$ (see Fig. \ref{fig:subfigures1d}). The contributions of both regions to (\ref{eq:EntSum}) are independent, and the integrand is smooth for $k>m$:

\begin{equation}\label{eq:cont1}
    \int_{\bm{k}\in \overline{A}_{m}}s_{2}^{\hat{a}+}(\bm{k})d^2k \sim f(\alpha)|m|^{2}
\end{equation}

\noindent
where we have dropped terms independent of $m$ and $f$ is a function of the polar angle $\alpha$ which defines $\hat{a}$. In the region $A_{m}$, $s_{2}^{\hat{a}+}(\bm{k})$ can be expanded asymptotically:

\begin{equation}\label{eq:cont2}
    \int_{\bm{k}\in A_{m}}s_{2}^{\hat{a}+}(\bm{k})d^2k \sim C_{0}(\alpha)\left(\frac{|m|}{k_{A}}\right)^{2} + C_{2}(\alpha)\left(\frac{|m|}{k_{A}}\right)^{2}\ln\left[\frac{|m|}{k_{A}}\right]
\end{equation}

Clearly, the contributions (\ref{eq:cont1}) and (\ref{eq:cont2}) are non-analytic functions of the mass $m$. The precise form of this non-analyticity depends on the choice of orbital $\hat{a}$, but the leading-order singular contribution is typically of the form $|m|^{2}\ln m$. Similar considerations for models with Dirac points in $D$ spatial dimensions also yield a leading order contribution to $s_{2}$ of the form $|m|^{D}\ln m$.

\begin{figure}%[H]
\begin{center}
\begin{tikzpicture}
    
    \draw[solid] (0,0) circle (2.2);
    \path [fill=green, fill opacity = 0.2] (0,0) circle (2.2);
    \path [fill=black, fill opacity = 1] (0,0) circle (0.1);
    \begin{scope}
        \path [fill=blue, fill opacity=0.2] (0,0) circle (1.0);
    \end{scope}
    \draw[thick,->] (0,0) -- (0,-2.2) node[anchor=south west] {\large{${k_A}$}};

    \draw[thick,->](0,0) -- (1,0);
    \node at (0.6,0.22) {\large{$m$}};
    
    \filldraw[black] (-2,0) circle (0pt);
    \node at (-1.5,0.05) {\large{$\overline{A}_{m}$}};
    \filldraw[black] (-1,0) circle (0pt) node[anchor=west] {\large{${A_{m}}$}};
\end{tikzpicture}
	\caption{The partition of momentum space near the Dirac point. The region $A_{m}$ contains momenta within a radius $m$ of the Dirac point at the origin. Its complement $\overline{A}_{m}$ contains momenta up to the scale $k_{A}$, which can be taken to be $1$.}
	\label{fig:subfigures1d}
	\end{center}
\end{figure}
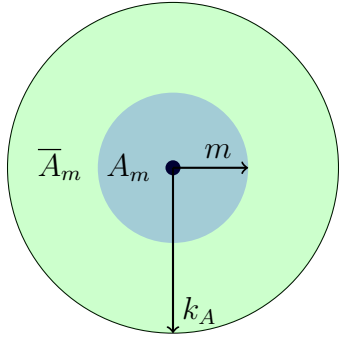

\end{widetext}

\end{document}